\documentclass{article}
\usepackage{hiph-preprint}
\usepackage{graphicx}
\volnumber{22} \issuenumber{1} \edyear{2005}                             
\frompage{000} \topage{000}                                              
\recrevdate{1 January 2005}                                              
\def\rightmark{Return of the Volcano}
\def\leftmark{McCumber, M. and Frantz, J.}

\newcommand{\sqrtsnn}{\mbox{$\sqrt{s_{NN}}$}}
\newcommand{\dphi}{\mbox{$\Delta\phi$}}

\title{Return of the Volcano:  PHENIX Azimuthal Correlations 62.4 GeV $Au+Au$
}
\authors{
{Michael McCumber and Justin Frantz %
\index{McCumber, M.} 
\index{Frantz, J.} 
}\\[2.812mm]
{\normalsize
\hspace*{-8pt}$^1$ SUNY Stony Brook Physics,\\
11794 Stony Brook, NY, USA\\[0.2ex]
}}

\abstract{ As in previous analyses at \sqrtsnn\ 200 GeV,
correlations in azimuthal angles between inclusive charge particles
at intermediate transverse momentum ($p_T$ = 1.0-4.0) GeV are
studied at \sqrtsnn\ 62.4 GeV.  The di-jet correlations reveal similar
modification as in 200 GeV. Specifically large modification,
including the "volcano" or "cone" structure, persists in the
awayside correlation.}


\PACS{see: http://www.aip.org/pacs/ }

\makeindex
\begin{document}

\maketitle

\section{Introduction}\label{intro}

Much effort has been made toward understanding the medium effects in
\sqrtsnn\ = 200 GeV Au+Au collisions at RHIC via jet properties
extracted from correlations in di-hadron azimuthal opening angle.
Initially these results included the observation of quenching,
\cite{initphenix},\cite{initstar} and later broadening in the very
low momentum ($<$ 1 GeV) (presumed) jet particles compared to such
correlations from normal jet fragmentation.
\cite{starmikemillerqm04}.  In \cite{ppg032}, PHENIX showed that
under 2-source (flow+jet) assumptions, the awayside correlations have a
distinctive "volcano-like" or "conical" structure, with a peak
displaced from 180 degrees from the nearside trigger particle, in
stark contrast to the normal "back to back" jet fragmentation
correlation which peaks at 180.  This interesting feature is also
evident in STAR data obtained with similar methods
\cite{starConfirmCone}, albeit with less significance due to a
larger systematic error assignment for the $v_2$ subtraction.
Nonetheless the observations have generated much speculation as to
possible physical "conical flow" origins of the displaced peak
\cite{colorCerenkov},\cite{machCone}, \cite{bendingetc}.  We extend
these works by applying a similar analysis to \sqrtsnn\ = 62.4 GeV
Au+Au collisions in order to observe whether such effects continue
to exist at this lower energy.

\section{Analysis}\label{ana}

\begin{figure}
\centering
\includegraphics[height=0.55\textheight]{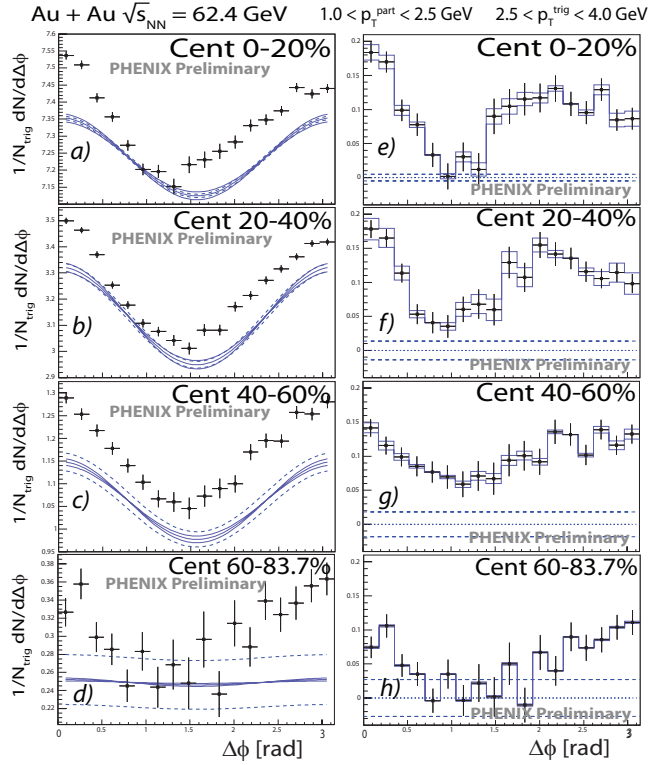}
\caption{\textit{a)-d)}: Per-trigger $dN/\Delta\phi$ yields without
combinatorial background subtraction, proportional to the two
particle $\Delta\phi$ correlation function (see text).  The
calculated background contribution is shown in the solid curves
[dashed curve 1-$\sigma$ uncertainty]. \textit{e)-h)}: Background
subtracted $dN_{dj}/\Delta\phi$ per-trigger conditional yields.  The
awayside "volcano" is evident in the more central bins.}
\label{fig:fgandbgphi}
\end{figure}

This analysis is based on a sample of 41.4 M minimum bias events
taken during the RHIC Run4 in 2004 at $\sqrt{s_{NN}}$ = 62.4 GeV. We
follow the basic methods outlined in \cite{ppg032},\cite{ppg033}.
Two categories of unidentified charged hadron tracks are considered
based on track $p_T$: \emph{trigger} ($t$) particles [2.5 GeV/c $<
p_T <$ 4.0 GeV/c] and \emph{partner} ($p$) particles [1.0 GeV/c $<
p_T <$ 2.5 GeV/c]. We define the number of di-jet correlated
particle pairs per event as a function of the \dphi\ between the two
particles, $dN_{dj}/d\Delta\phi$, according to the formula

\begin{equation}
\frac{1}{N_{evts}} \frac{dN_{dj}}{d\Delta\phi} =
\frac{1}{N_{evts}Acc(\Delta\phi)} \left(
\frac{dN_{pairs}}{d\Delta\phi} - \frac{dN_{bkg}}{d\Delta\phi}
\right)
\label{eq:full}
\end{equation}

where $dN_{pairs}/d\Delta\phi$ represents all $t$-$p$ pairs made in
real events and $Acc(\Delta\phi)$ represents the pair acceptance and
efficiency correction function. The $\Delta\phi$ shape of
 $Acc(\Delta\phi)$ is derived from a mixed event procedure,
making pair combinations taking $t$ and $p$ particles from different
events, and its overall normalization is set by the single
particle acceptance and efficiency in the $t$/$p$ category $p_T$
ranges. $dN_{bkg}/d\Delta\phi$ is generated by the exact same event
mixing procedure, therefore dividing it by $Acc(\Delta\phi)$ results
in a constant function of $\Delta\phi$.  Under the assumptions of
the 2-source model \cite{ppg032}, this constant background is
modulated by the independently determined per-event averaged flow
measurements, $\langle v^t_2 \rangle$ and $\langle v^p_2 \rangle$ according to:

\begin{equation}
\frac{1}{N_{evts}Acc(\Delta\phi)} \frac{dN_{bkg}}{d\Delta\phi} = B
 \left( 1 + 2\langle v^{t}_{2}\rangle\langle v^{p}_{2} \rangle
\cos(2\Delta\phi) \right) \label{eq:bkg}
\end{equation}

where $B$ defines the normalization of the non-interesting
background.  To obtain $B$ in this analysis we employ an "absolute"
method also used in \cite{ppg033}, as well as the "Zero Yield At
Minimum" (ZYAM) method described in \cite{ppg032} for comparison. In
the absolute method, it is assumed the background level $B$ is
simply the product of the per-event averaged single's multiplicities
for both the trigger and partner particle categories, along with an
additional factor $\xi$ which accounts for residual multiplicity
correlations: that is, $B = \xi\langle n_t \rangle \langle n_p
\rangle$. Neglecting $\xi$, this assumption has a first-order error
equal to $\langle j_t\rangle \langle j_p \rangle$, where $j_x$
represents the number of true jet particles in category $x$. In
$p+p$ events it was checked that compared to the
signal size $\langle n_tn_p \rangle
\approx \langle j_tj_p \rangle$, $\langle j_t \rangle \langle j_p \rangle$ is
negligible, and it is assumed that in $Au+Au$, where the background
and combinatoric contributions dominate, the approximation
continues to hold.  The need for $\xi$, a correction for the finite
sized centrality binning/resolution effects in the mixing
discussed in \cite{ppg033}, is due to residual multiplicity
correlation between numbers of jet and non-jet particles due to
event geometry/global event multiplicity, which affects the overall
normalization but not the $\Delta \phi$ shape of the total
correlation.  We scale our final results per trigger particle.

\section{Results, Discussions, and Conclusions}\label{others}

In figure \ref{fig:fgandbgphi} the shape of the correlations are
shown for 4 centrality bins.
Since $Acc(\Delta\phi)$ as calculated is proportional to the product of
the independent $t$/$p$ probabilities $P(t)$ and $P(p)$, the quantity
in the left column \textit{a)-d)}, $1/Acc(\Delta\phi)(dN_{pairs}/d\Delta\phi)$,
is proportional to the raw correlation strength
$P(t,p)/P(t)P(p)$.  Figure \ref{fig:fgandbgphi} \textit{e)-h)}
are the $dN_{dj}/d\Delta\phi$ yields, eq. (\ref{eq:full}), after subtraction
of the harmonic background.  Both a gaussian shaped peak on the nearside (and
awayside in the two peripheral selections), typical of jet
correlations in this kinematic regime \cite{ppg032}, as well as the
displaced "volcano" peak in the awayside region previously observed
at 200 GeV are evident.

\begin{figure}
\centering
\includegraphics[width=1.0\textwidth]{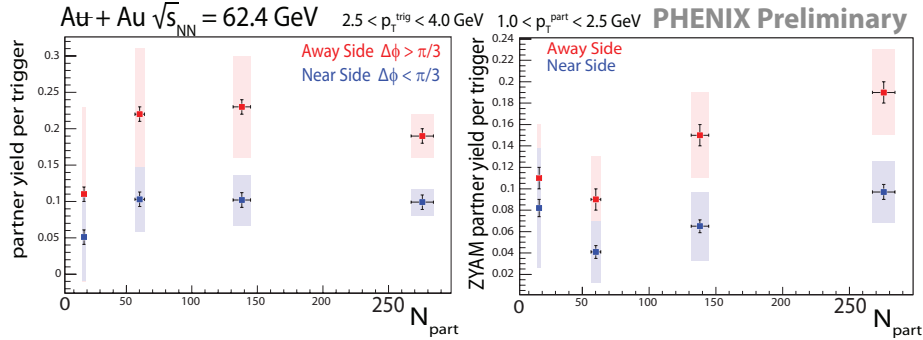}
\caption{Nearside and awayside integrated per-trigger conditional
yields $a)$ using the "absolute" background normalization method and
$b)$ using the ZYAM method, both $vs.$ number of participants.}
\label{fig:yields}
\end{figure}

In figure \ref{fig:yields}, the yields are integrated
and plotted vs. the collision centrality (parameterized as
$N_{part}$).  Within large uncertainties, no clear centrality
dependence as seen in \cite{ppg032} can be confirmed or denied,
though the comparison between these integrated yield results and the
200 GeV \cite{ppg032} (not shown) shows that the 62.4 GeV near-side
yields are lower, with significance, for all centrality bins.  This
behavior is expected for jet fragmentation, due to a larger
(high-$z$) trigger bias at lower energy due to the more steeply
falling hard-scattering cross sections.

In summary, the same abnormal awayside features of the apparent
di-jet correlations seen at 200 GeV are present in the 62.4 GeV
PHENIX data.  Combining information of the full energy dependence of
the modification along with possibly the dependence on system size
as in such studies of Cu+Cu collisions should constrain the possible
theoretical explanations of their origins.

\vfill\eject
\end{document}